# First results from the HENSA/ANAIS collaboration at the Canfranc Underground Laboratory

N Mont-Geli[1], A Tarifeño-Saldivia[1], S E A Orrigo[2], J L Taín[2], M Grieger[4], J Agramunt[2], A Algora[2], J Amaré[6], D Bemmerer[4], F Calviño[1], S Cebrián[6], I Coarasa[6], G Cortés[1], A De Blas[1], I Dillmann[5], L M Fraile[3], E García[6], R García[1], M Martínez[6], E Nacher[2], Y Ortigoza[6], A Ortiz[6], M Pallàs[1], J Puimedón[6], A Salinas[6], M L Sarsa[6] and A Tolosa-Delgado[2]

[1] Institut de Tècniques Energètiques, Univ. Politècnica de Catalunya, Barcelona, Spain
[2] Insituto de Física Corpuscular, CSIC - Univ. de Valencia, Valencia, Spain
[3] Grupo de Física Nuclear & IPARCOS, Univ. Complutense de Madrid, Madrid, Spain
[4] Helmholtz-Zentrum Dresden-Rossendorf, Dresden, Germany
[5] TRIUMF, Vancouver, Canada
[6] Centro de Astropartículas y Física de Altas Energías, Universidad de Zaragoza, Zaragoza, Spain

nil.mont@upc.edu

**Abstract**. The HENSA/ANAIS collaboration aims for the precise determination of the neutron flux that could affect ANAIS-112, an experiment looking for the dark matter annual modulation using NaI(Tl) scintillators. In this work, the first measurements of the neutron flux and Monte Carlo simulations of the neutron spectrum are reported.

## 1. Introduction

The HENSA/ANAIS collaboration aims for the precise determination of the neutron flux at hall B of the Laboratorio Subterráneo de Canfranc (LSC) [1], that could affect ANAIS-112, an experiment looking for the dark matter annual modulation using NaI(Tl) scintillators [2].

The High Efficiency Neutron Spectrometry Array (HENSA) [3] is a detector system based on the Bonner Spheres principle [4]. In order to be sensitive at different energy ranges, HENSA is composed by up to ten $^3$He-filled proportional counters embedded in High Density PolyEthylene (HDPE) moderators with different sizes. The neutron spectrum is unfolded from the counting rates and the response matrix by means of appropriate reconstruction algorithms [5]. Key to reliability of the reconstruction process is the use of a priori information [6]. This is typically achieved by means of realistic Monte Carlo simulations or from previous measurements [7].

Early versions of HENSA have already been used for the assessment of the neutron flux in hall A of the LSC [8] and in the shallow underground facility Felsenkeller in Dresden [9]. Currently, a long-term characterization of the neutron flux is being carried out at the LSC: in hall A from October 2019 up to March 2021 [10] and in hall B since March 2021.

This work focuses on measurements of the neutron flux in hall B and Monte Carlo simulations of the neutron spectrum in hall A.

## 2. Neutron flux measurements in hall B

Measurements started in March 2021 and a long-term characterization is foreseen, at least, until December 2022. Currently, only three HENSA detection modules are being used, however the full setup is expected to be running by December 2021. As it is shown in figure 1, the current HENSA setup (detectors 1, 2 and 5), covers the thermal, epithermal and evaporation regions of the spectrum.

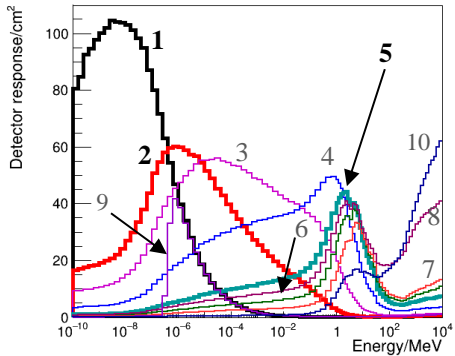

**Figure 1.** HENSA response matrix. The modules which are currently being used are shown with a thicker line (1, 2 and 5).

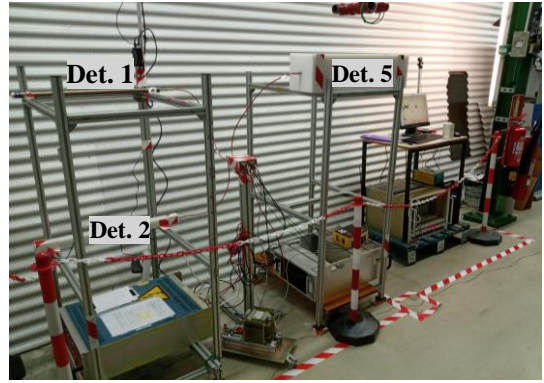

**Figure 2.** HENSA setup in hall B, close to ANAIS-112. Currently, only three detection modules are being used. The full setup is expected to be running by December 2021.

### 2.1. Data analysis

Figure 3 shows an energy deposition ($E$) spectrum measured with a HENSA proportional counter (red region). Due to the low counting rates, the contributions of γ rays, α particles, electronic noise and neutrons are overlaid. The high-energy component (green region) is due to the α-radioactivity coming from the counter walls [11] and is assumed to be linear. The neutron component is determined as,

$$S'_B(E) = S_B(E) \cdot \frac{A_R - \alpha}{A_B} = S_B(E) \cdot f \quad (1)$$

In the previous equation $S_B(E)$ is a reference neutron energy deposition spectrum obtained from measurements using a $^{252}$Cf source. The multiplying factor $f$ is calculated inside the 700 to 800 keV region, the so-called full-energy deposition peak, using the area under the experimental spectrum ($A_R$), the area under the reference spectrum ($A_B$), and the alpha particles component ($\alpha$). The total number of detected neutrons is, therefore, determined as the area of the neutron component ($S'_B(E)$).

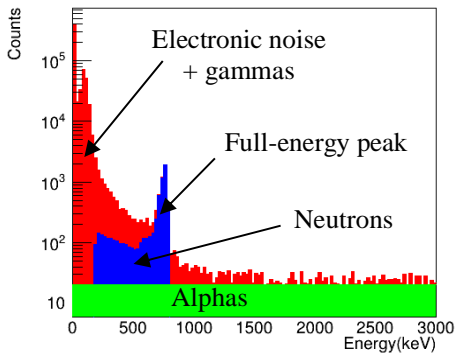

**Figure 3.** Energy deposition spectrum. The neutron component is shown in blue while the alpha particles component is shown in green.

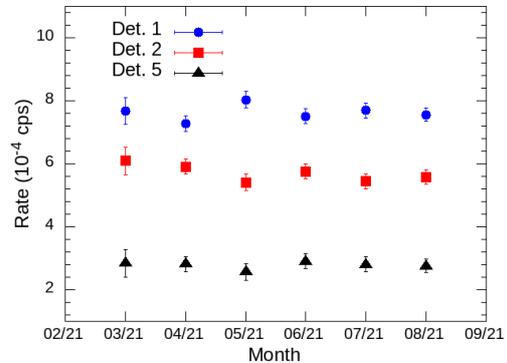

**Figure 4.** Neutron counting rates in hall B: blue circles correspond to detector 1, red squares to detector 2 and black triangles to detector 5.

*2.2. Results*

As it is shown in figure 4, the neutron counting rates are stable. However, the measurement period is not large enough to get a definitive conclusion about this topic. Additionally, it has to be considered that with the current setup we are not able to cover the totality of the neutron spectrum.

The average neutron counting rates are $8(3) \cdot 10^{-4}$ counts per second (cps) for detector 1, $5.9(3) \cdot 10^{-4}$ cps for detector 2 and $3.1(3) \cdot 10^{-4}$ cps for detector 5. These values can be compared with the rates measured in hall A using a similar setup: $4.64(23) \cdot 10^{-4}$ cps for detector 1 and $4.37(15) \cdot 10^{-4}$ cps for detector 2. The difference could be explained because in hall B detectors are closer to the walls than in hall A.

**3. Simulation of the neutron spectrum in hall A**

*3.1. Simulation setup*

It is well known that the main sources of neutrons in deep underground laboratories are (α,n) reactions and spontaneous fission processes due to the U and Th natural decay chains [12]. Previous calculations have shown that at the LSC (α,n) reactions are dominant [13]. Therefore, the key inputs for the simulation are the chemical composition and the intrinsic radioactivity of the α particle sources: the mountain rock and the concrete walls of the laboratory.

A detailed characterization of the rock [13] has shown a great heterogeneity so that calculations based on different samples (the so-called T14, T31 and T51) have been carried out. For the laboratory walls, several types of standard concretes [14] have been used. Nevertheless, a future detailed characterization of their chemical composition is expected. Information about the concrete radioactivity has been provided by the LSC staff.

Calculations of the neutron transport have been carried out using FLUKA [15,16]. The α-induced neutron emission rate and spectrum in both the rock and the concrete are calculated using NeuCBOT [17] and transformed into a FLUKA input using the same methodology as Grieger *et al.* [9].

*3.2. Results*

Figure 5 shows the neutron flux spectrum due to (α,n) reactions inside several types of rock (a) and concrete (b). The most relevant result is the fact that (α,n) reactions inside the concrete walls are the main contribution. The shape of the energy spectrum is also determined by the type of concrete. Therefore, a detailed characterization of its chemical composition and density is mandatory for a deeper understanding of the neutron source.

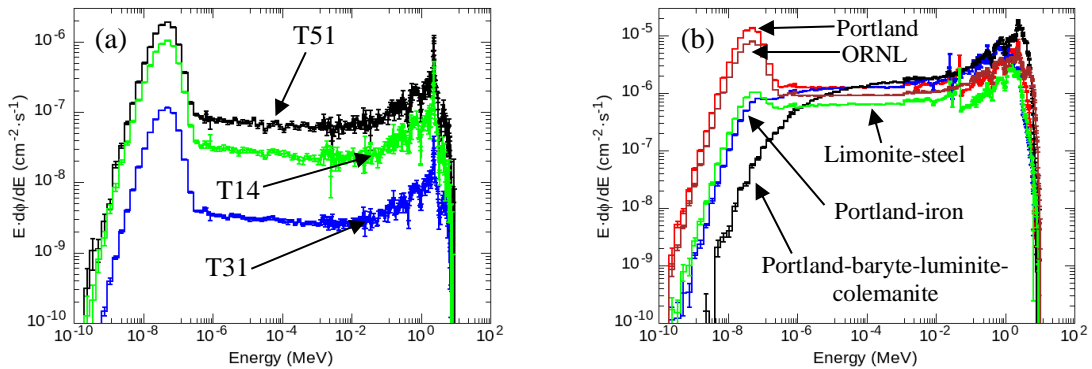

**Figure 5.** Neutron flux spectrum due to (α,n) reactions inside the rock (a) and inside the concrete walls (b). See section 3.1. for more information. ORNL indicates the concrete used in the Oak Ridge National Laboratory (Oak Ridge, USA).

The neutron flux measured in 2013 (empty hall A) was $(13.8 \pm 1.4) \cdot 10^{-6}$ cm$^{-2} \cdot$ s$^{-1}$ [7]. This value can be compared with the calculated fluxes which are reported in table 1. The overestimation of the absolute magnitude of the neutron flux might be explained due to and overestimation of the α-induced neutron yields when using NeuCBOT [18] and due to the uncertainties of the (α,n) cross-sections.

**Table 1.** Integrated neutron flux (cm$^{-2}$ · s$^{-1}$) due to (α,n) reactions inside the rock and the concrete walls.

| Type | Concrete flux | Density (g/cm$^3$) | Sample | Rock flux | Activity (Bq/kg) [13] |
|---|---|---|---|---|---|
| Portland | 63.22(6) · 10$^{-6}$ | 2.3 | T51 | 4.61(2) · 10$^{-6}$ | 107(2) |
| ORNL | 45.03(3) · 10$^{-6}$ | 2.3 | T14 | 2.48(1) · 10$^{-6}$ | 90(7) |
| Portland - Baryte | 47.07(7) · 10$^{-6}$ | 3.1 | T31 | 0.2769(8) · 10$^{-6}$ | 13.0(4) |
| Limonite - Steel | 17.81(3) · 10$^{-6}$ | 4.5 | | | |
| Portland - Iron | 35.42(4) · 10$^{-6}$ | 5.9 | | | |

## 4. Summary and future work

It has been shown that the neutron counting rates are stable. However, the measurement period is not large enough to get a definitive conclusion. Another interesting result is the fact that counting rates in hall B are greater than the ones measured in hall A with a similar setup. The difference could be explained because in hall B detectors are closer to the walls than in hall A. Future Monte Carlo simulations are expected to provide more information about this topic. Additionally, an improvement of the data analysis method is expected by means of pulse shape and machine learning techniques.

Calculations seem to overestimate the absolute magnitude of the neutron flux. It might be explained due to and overestimation of the α-induced neutron yields when using NeuCBOT and due to the uncertainties of the (α,n) cross-sections. For the near future, we plan to assess the impact of using different data libraries and calculation tools. Additionally, a detailed study about the effects of the air humidity, the rock and concrete humidity and the position of HENSA inside the experimental hall is expected to be carried out. Nevertheless, any attempt to improve the simulation requires a characterization of the chemical composition and density of the concrete walls.

## 5. Acknowledgements

This work has been supported by the Spanish Government funds FPA2017-83946-C2-1 & C2-2, PID2019-104714GB-C2-1 & C2-2, PID2019-104374GB-I00 and RTI2018-098868-B-I00. We also acknowledge the Generalitat Valenciana Grant No. PROMETEO/2019/007.